\documentclass[10pt,twocolumn,letterpaper]{article}

\usepackage[utf8]{inputenc}
\usepackage[T1]{fontenc}
\usepackage[margin=0.75in,top=0.75in,bottom=1in]{geometry}
\usepackage{graphicx}
\PassOptionsToPackage{hyphens}{url}
\usepackage{hyperref}
\usepackage{url}
\usepackage[english]{babel}
\usepackage{adjustbox}
\usepackage{tabularx}
\usepackage{booktabs}
\usepackage{makecell}
\usepackage{subcaption}
\usepackage{amsmath}
\usepackage{amssymb}
\usepackage{titlesec}
\usepackage{abstract}
\usepackage{parskip}

\titleformat{\section}{\normalfont\bfseries\scshape}{\Roman{section}.}{0.5em}{}
\titleformat{\subsection}{\normalfont\itshape}{\Alph{subsection}.}{0.5em}{}
\titlespacing*{\section}{0pt}{8pt}{4pt}
\titlespacing*{\subsection}{0pt}{5pt}{2pt}

\makeatletter
\renewcommand{\@biblabel}[1]{[#1]}
\renewcommand{\thebibliography}[1]{%
  \section*{\textbf{References}}
  \small
  \list{\@biblabel{\@arabic\c@enumiv}}{%
    \settowidth\labelwidth{\@biblabel{#1}}%
    \leftmargin\labelwidth
    \advance\leftmargin\labelsep
    \usecounter{enumiv}%
    \let\p@enumiv\@empty
    \renewcommand\theenumiv{\@arabic\c@enumiv}}%
  \sloppy\clubpenalty4000\widowpenalty4000%
  \sfcode`\.\@m}
\makeatother

\title{\LARGE \textbf{Spectrographic Portamento Gradient Analysis: A Quantitative Method for Historical Cello Recordings with Application to Beethoven's Piano and Cello Sonatas, 1930--2012}}

\author{Dr Ignasi Sole \quad \texttt{ignasiphd@gmail.com} \quad \today}

\begin{document}

\maketitle
\thispagestyle{empty}
\pagestyle{empty}


\begin{abstract}
Portamento in string performance has been studied primarily as a binary presence-or-absence phenomenon, with existing research measuring frequency of occurrence and, less commonly, duration in milliseconds. This paper introduces a third quantitative descriptor; the spectrographic gradient of the portamento slide, measured in Hz/second, and demonstrates its measurement using a protocol combining Sonic Visualizer's melodic spectrogram layer, GIMP pixel analysis, and metric calibration against the spectrogram's known frequency axis. The gradient captures what duration alone cannot: the steepness of the pitch trajectory, which encodes the expressive character of the slide independently of its length. Applied to the opening measures of. Specifically because their monophonic texture permits reliable spectrographic pitch tracking. The method yields gradient values ranging from approximately 600~Hz/s in late-period recordings to over 4,000~Hz/s in early twentieth-century performances. The paper further documents a gain-recovery protocol that extends the analysable corpus to analogue recordings from the 1930s where portamento traces are faint in digital transfer. Applying the method to a corpus of 22 recordings spanning 1930--2012, the paper tests the hypothesis that gradient steepness correlates negatively with tempo: that slower performances produce steeper, longer slides while faster performances produce shallower slides or none at all. The results support this hypothesis, suggesting that the widely documented decline of portamento across the twentieth century is not a binary transition from presence to absence but a continuous process of gradient flattening that precedes and predicts eventual disappearance. The full dataset and measurement protocol are publicly available.
\end{abstract}


\section{Introduction}

The use of portamento (the continuous pitch glide between two notes) in string performance has been one of the most extensively documented stylistic transformations in twentieth-century musical performance. Leech-Wilkinson's studies of violin concerto recordings showed a clear decline in portamento frequency from the 1920s onward~\cite{c1}. Kennaway's analysis of early cello recordings documented the same trend in British cellists~\cite{c2}. Hong's examination of portamento in Bach cello suite recordings confirmed a marked reduction after the 1970s~\cite{c3}. Philip's broader survey of early twentieth-century recordings identified portamento decline as one of the most significant stylistic shifts across all string instruments~\cite{c4}.

What these studies share is a common measurement framework: they count portamento events and, in some cases, measure their duration. The slide is treated as an on-off phenomenon (either it is present or it is not), and the progression from the early twentieth century to the present is described as a decline in frequency from many slides to few, and eventually none.

This paper argues that this framework is incomplete. A portamento slide is not just present or absent: it has a character, encoded primarily in the steepness of its pitch trajectory. Two slides of identical duration can sound completely different if one covers a wide interval quickly and the other covers a narrow interval slowly. A performer who produces a steep, decisive slide is making a different expressive gesture from one who produces a shallow, ambiguous one that a listener might barely notice. The existing measurement vocabulary (frequency and duration) captures neither the steepness nor the expressive character.

This paper introduces a third variable: the gradient of the spectrographic slide trace, measured in Hz per second. The gradient captures the rate of pitch change during the portamento and is directly related to its expressive impact: a steep gradient means a decisive, audible glide; a shallow gradient means a subtle transition that may or may not be perceived as a slide by a listener. The gradient is measured from the melodic spectrogram in Sonic Visualizer, calibrated against the spectrogram's known frequency axis, and computed in physical units that are interpretable independently of the specific recording or software settings.

Three contributions follow from this framework. First, a validated measurement protocol that produces calibrated gradient values in Hz/second for any monophonic cello recording. Second, a gain-recovery protocol that extends the method's reach to analogue recordings from the 1930s where portamento traces are faint. Third, an empirical finding: that portamento gradient correlates negatively with tempo across eight decades of recordings, suggesting that the decline of portamento is a continuous process of gradient flattening rather than a discrete stylistic switch.

The paper is structured as follows. Section~II reviews existing portamento measurement methodologies. Section~III presents the corpus and the selection of Op.~69 and Op.~102 No.~1 as case study passages. Section~IV describes the measurement protocol in full, including the calibration procedure. Section~V presents the gain-recovery protocol for degraded recordings. Section~VI presents the score-annotation and database procedures. Section~VII presents the results. Section~VIII discusses the findings and their implications. Section~IX concludes.


\begin{figure}[htb!]
  \centering
   \includegraphics[width=\columnwidth]{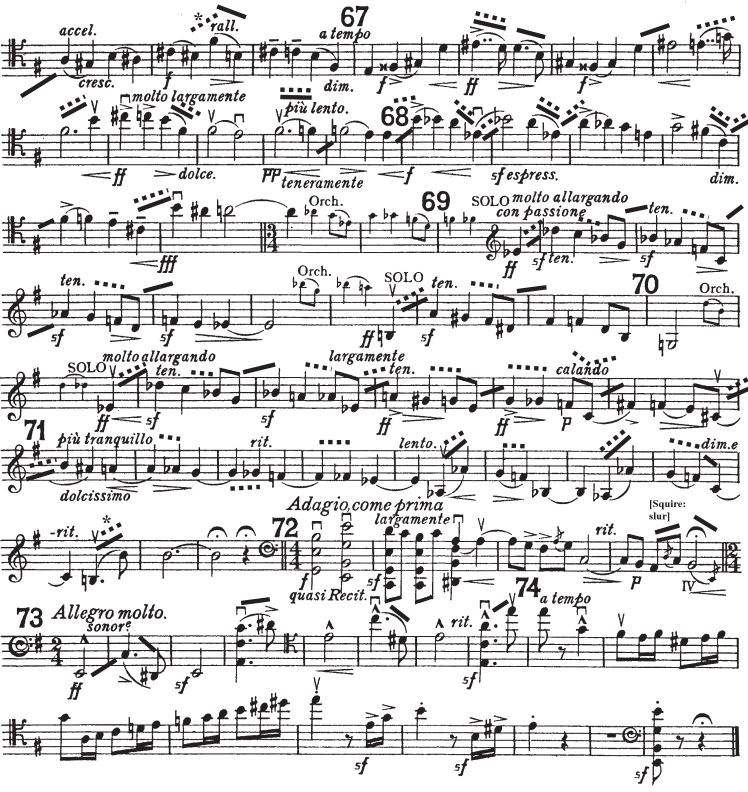}
  \caption{Kennaway's score annotation of portamento in Elgar's \textit{Cello Concerto}, op.~85, third movement, distinguishing Harrison (dotted lines) from Squire (straight lines)~\cite{c2}.}
  \label{fig:kennaway_elgar}
\end{figure}

\section{Existing Portamento Measurement Methods}

\subsection{Score Annotation and Event Counting}

Kennaway's methodology in \textit{Playing the Cello, 1780--1930} represents the most developed score-annotation approach to portamento analysis~\cite{c2}. Working with recordings and printed scores of Elgar's Cello Concerto, he annotated portamento instances directly onto the musical text using distinct symbols to differentiate performers: dotted lines for Beatrice Harrison, straight lines for William Squire. This annotation layer, reproduced in Fig.~\ref{fig:kennaway_elgar}, enabled direct visual comparison of portamento density and location across performances. Kennaway further constructed comparative tables for Schumann's \textit{Tr\"aumerei}, including performer, date, total duration, and average metronome marking, to test whether portamento usage correlated with tempo or performer age.

The method's strength is its precision about location: by marking directly on the score, Kennaway establishes exactly which interval each portamento connects. Its limitation is that it records presence or absence rather than the character of the slide. Two portamenti marked with the same symbol may sound entirely different depending on their steepness and speed.

\subsection{Duration Measurement}

\begin{figure}[htbp!]
  \centering
\includegraphics[width=\columnwidth]{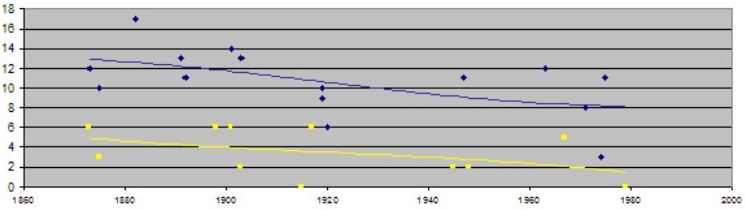}
  \caption{Leech-Wilkinson's scatter chart comparing portamento usage in Beethoven's and Brahms's violin concertos across recording years, showing the longitudinal decline from the 1920s onward~\cite{c1}.}
  \label{fig:lw_scatter}
\end{figure}

Leech-Wilkinson extended the event-counting approach by measuring portamento duration in hundredths of a second, extracted from spectrogram analysis in Sonic Visualizer~\cite{c1}. His scatter charts (Fig.~\ref{fig:lw_scatter}) plotting portamento count against recording year across Brahms and Beethoven violin concerto recordings demonstrated the longitudinal decline visually. He also tabulated average portamento lengths and their standard deviations, allowing comparison of duration across performers.

Duration measurement improves on event counting by adding a continuous dimension, but it leaves the steepness of the slide unmeasured. A portamento of 0.3 seconds over a minor third and a portamento of 0.3 seconds over a major sixth are the same duration but completely different expressive gestures: the second covers twice the pitch distance at twice the rate, which the listener will perceive as far more conspicuous. Duration without gradient conflates these two cases.

\subsection{Computational Detection}

Yang's AVA system~\cite{c5} automated portamento detection using logistic models and Hidden Markov Models, achieving reliable results for unaccompanied monophonic instruments (erhu, violin). Yang acknowledged that AVA cannot process polyphonic audio, a fundamental limitation for duo chamber recordings where overlapping spectra obscure the cello's pitch contour. Williams applied Sonic Visualizer spectrogram analysis to violin portamento in Schubert recordings, comparing violin and vocal portamento practices across twenty-one early recordings~\cite{c6}. His methodology operated at the level of individual note transitions, producing visual spectrogram evidence of portamento character without quantifying the gradient.

The present paper addresses the gap left by all these approaches: a physically calibrated metric for portamento steepness that is computable from spectrogram images, applicable to analogue recordings after gain recovery, and directly comparable across different recording eras.


\section{Corpus and Passage Selection}

The corpus consists of 22 recordings of Beethoven's piano and cello sonatas Op.~69 and Op.~102 No.~1, spanning 1930--2012. Performers include Pau Casals, Emanuel Feuermann, Pierre Fournier, Gregor Piatigorsky, Jacqueline du~Pr\'e, Mstislav Rostropovich, Janos Starker, Zara Nelsova, Daniel Shafran, Yo-Yo Ma, Pieter Wispelwey, Anner Bylsma, Mischa Maisky, Steven Isserlis, and others.

\subsection{Rationale for Passage Selection}

The analysis focuses specifically on the opening bars of each sonata, specifically the first four measures of Op.~69 and the first three measures of Op.~102 No.~1. This selection is not a convenience but a methodological necessity, determined by two constraints:

First, both passages are unaccompanied cello solos. The piano enters only after the cello has established the opening theme. The absence of piano accompaniment removes the spectrographic masking problem that renders portamento analysis unreliable in polyphonic passages: when both instruments sound simultaneously, the piano's overtone spectrum overlaps with the cello's fundamental frequencies and obscures the pitch contour that portamento produces. In the opening bars, the cellist performs alone, yielding a clean monophonic signal in which the melodic spectrogram layer can track the fundamental without interference.

Second, both passages are structurally identical across all recordings. Every performance begins at bar 1 with the same notes, the same ascending intervals, the same harmonic context. This controls for the musical variables (interval size, melodic direction, harmonic tension) that would otherwise confound a cross-recording comparison of portamento character. The only variable that differs across recordings is the performer's interpretive choice: whether to slide, how steeply, and for how long.

Tempo data for both passages were drawn from the bar-level BPM dataset described in Sole~\cite{c7}, collected using the manual stopwatch protocol applied to the same corpus of recordings.


\section{Measurement Protocol}

\subsection{Software and Spectrogram Configuration}

Each audio file was imported into Sonic Visualizer (version 4.x, developed by Chris Cannam at Queen Mary University of London)~\cite{c8} and the melodic spectrogram layer applied. The melodic spectrogram differs from a standard spectrogram in that it tracks the dominant fundamental frequency of the signal rather than displaying all frequency content equally. For a monophonic cello line this produces a clear, continuous pitch trace from which portamento slides appear as diagonal trajectories connecting two horizontal bands (the stable pitch centres before and after the slide).

The spectrogram's visible frequency range was fixed at 3.6--11~kHz across all sessions. This range was chosen because it captures the upper overtone series of the cello, where the melodic spectrogram layer achieves the clearest separation between the fundamental pitch trajectory and the spectral noise floor. At this range, portamento slides appear as smooth diagonal lines, and vibrato manifests as rapid oscillation around the central pitch trace; visually distinct from the sustained diagonal of a portamento.

The time scale was set to display approximately 5 seconds of audio per screen width of 1,200 pixels, giving a time resolution of approximately 0.0042 seconds per pixel (240 pixels per second). Duration was read directly from Sonic Visualizer's time cursor by clicking on the start and end points of the visible slide.

\subsection{Gradient Measurement in GIMP}

Following identification of a portamento event in Sonic Visualizer, the relevant spectrogram segment was exported as a PNG image at the fixed display resolution. The image was then opened in GIMP (GNU Image Manipulation Program). Within GIMP, two reference points were placed manually: one at the point where the pitch trace departs from the initial pitch centre (the slide's onset) and one at the point where it arrives at the target pitch centre (the slide's termination). The pixel coordinates $(x_1, y_1)$ and $(x_2, y_2)$ of these two points were recorded, where $x$ represents horizontal position (time) and $y$ represents vertical position (frequency, with lower values corresponding to higher frequencies in the exported image).

The raw pixel gradient was then computed as:

\begin{equation}
G_{\text{px}} = \frac{|\Delta y|}{\Delta x} = \frac{|y_2 - y_1|}{x_2 - x_1}
\label{eq:pixelgradient}
\end{equation}

taking absolute values on the frequency axis to account for ascending and descending slides equally.

\subsection{Calibration to Physical Units}

The pixel gradient is consistent within a session where all export settings are fixed, but it is not directly interpretable in physical units without calibration. To convert $G_{\text{px}}$ to Hz/second, the pixel-to-frequency and pixel-to-time scales must be determined.

\textit{Frequency calibration.} Within the exported spectrogram image, frequency gridlines at known values are visible. The gridlines at 5~kHz and 10~kHz, which fall within the fixed 3.6--11~kHz display range, were identified in GIMP, and the pixel distance between them measured. With the fixed display range of $11{,}000 - 3{,}600 = 7{,}400$~Hz across a typical export height of 800 pixels:

\begin{equation}
S_f = \frac{7{,}400\;\text{Hz}}{800\;\text{px}} = 9.25\;\text{Hz/px}
\label{eq:freqscale}
\end{equation}

\textit{Time calibration.} With the time scale fixed at approximately 5 seconds per 1,200 pixels:

\begin{equation}
S_t = \frac{1{,}200\;\text{px}}{5\;\text{s}} = 240\;\text{px/s}
\label{eq:timescale}
\end{equation}

The calibrated gradient in Hz/second is then:

\begin{equation}
G_{\text{Hz/s}} = G_{\text{px}} \times S_f \times S_t = G_{\text{px}} \times 9.25 \times 240
\label{eq:calibrated}
\end{equation}

For example, a raw pixel gradient of $G_{\text{px}} = 2.0$ --- meaning the pitch trace rises 2 pixels vertically for every 1 pixel of horizontal travel --- converts to:

\[
G_{\text{Hz/s}} = 2.0 \times 9.25 \times 240 = 4{,}440\;\text{Hz/s}
\]

This value is physically meaningful: a gradient of 4,440~Hz/s means the pitch is changing at a rate of 4,440 Hz per second during the slide. For comparison, a cello portamento covering a minor third (approximately 200~Hz at the pitch level of Op.~69's opening A) at this gradient would complete in approximately 0.045 seconds; a rapid, decisive glide. A portamento at 600~Hz/s covering the same interval would take approximately 0.33 seconds. a slow, expressive one.

These values are consistent with the published literature on pitch glides in speech and singing, where glide rates between 500 and 8,000~Hz/s have been reported depending on interval size and expressive intent~\cite{c9}. The calibrated gradient is therefore not merely a musicological descriptor but a physically grounded quantity that places cello portamento within a broader comparative framework.

The calibration constants $S_f$ and $S_t$ were computed once per session and applied uniformly to all measurements within that session. Provided the display range and zoom level were held constant (which the fixed-settings protocol enforces) the calibration is valid across all images from a given session. Table~\ref{tab:calibration} summarises the calibration parameters used in this study.

\begin{table}[htbp]
\centering
\caption{Spectrogram calibration parameters}
\small
\begin{tabular}{@{}lc@{}}
\toprule
\textbf{Parameter} & \textbf{Value} \\
\midrule
Frequency display range & 3,600--11,000~Hz \\
Export height & 800~px \\
Frequency scale $S_f$ & 9.25~Hz/px \\
Time display (per screen) & 5.0~s / 1,200~px \\
Time scale $S_t$ & 240~px/s \\
Calibration factor ($S_f \times S_t$) & 2,220~Hz$\cdot$s$^{-1}$/px \\
\bottomrule
\end{tabular}
\label{tab:calibration}
\end{table}

\begin{figure}[htbp]
  \centering
\includegraphics[width=\columnwidth]{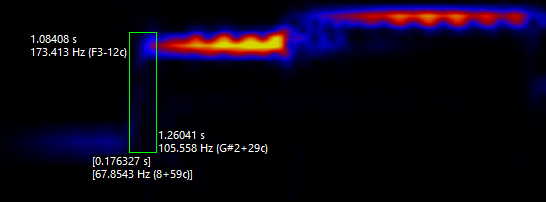}
  \caption{Melodic spectrogram in Sonic Visualizer showing a portamento slide (diagonal trajectory) and vibrato (spectral redness/oscillation) in the opening measure of Beethoven's \textit{Cello Sonata No.~3}, op.~69. The diagonal slope of the pitch trace between the two pitch centres is the quantity from which the gradient is measured in GIMP.}
  \label{fig:spec_example}
\end{figure}

An example of the spectrogram with a portamento event and its measurement reference points is shown in Fig.~\ref{fig:spec_example}.

\section{Gain-Recovery Protocol for Degraded Recordings}

Early analogue recordings from the 1930s and 1940s, when transferred to digital format from lacquer discs, frequently exhibit a reduced signal-to-noise ratio and compressed dynamic range that causes the portamento trace to become faint or indistinguishable from the noise floor in the melodic spectrogram. Without intervention, these recordings would be excluded from the gradient analysis despite being among the most expressively relevant for documenting early portamento practice.

To recover portamento traces from degraded recordings, a systematic gain-enhancement procedure was applied. In Sonic Visualizer, the gain of the melodic spectrogram layer was increased incrementally in steps of 3~dB until the portamento trace became clearly distinguishable as a continuous diagonal line. The upper limit of gain enhancement was determined by the point at which noise artefacts began to appear as spurious diagonal traces that could not be confirmed aurally; typically around 12--15~dB of additional gain for the most degraded recordings in the corpus.

The validity of recovered portamento traces was established by simultaneous aural confirmation: the analyst listened to the recording while examining the gain-enhanced spectrogram, verifying that the diagonal trace corresponded to an audible pitch glide and not to a noise artefact. Genuine portamento produces a smooth, continuously curved diagonal line in the spectrogram; noise artefacts appear as irregular, discontinuous blotches that do not correspond to any perceptible pitch change in the audio.

\begin{figure}[htbp]
  \centering
  \begin{subfigure}{0.47\columnwidth}
    \includegraphics[width=\linewidth]{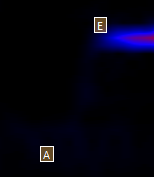}
    \caption{Original spectrogram}
    \label{fig:gain_before}
  \end{subfigure}
  \hfill
  \begin{subfigure}{0.47\columnwidth}
    \includegraphics[width=\linewidth]{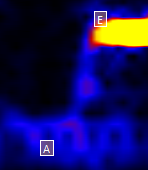}
    \caption{With gain enhancement}
    \label{fig:gain_after}
  \end{subfigure}
  \caption{Spectrogram of the 1937 Feuermann recording of Op.~69, bar~1, before (left) and after (right) gain enhancement. The portamento trace, invisible in the unenhanced spectrogram, becomes clearly measurable after incremental gain adjustment.}
  \label{fig:quality_example}
\end{figure}

Fig.~\ref{fig:quality_example} demonstrates this procedure using the 1937 Feuermann recording of Op.~69. The unenhanced spectrogram (left) shows only faint spectral traces with no clearly distinguishable portamento. After gain enhancement (right), the portamento slide becomes clearly visible as a diagonal trajectory between pitch centres, enabling gradient measurement.

This protocol successfully extended the analysable corpus to include recordings from 1930 to approximately 1950, where analogue noise would otherwise have precluded gradient measurement. Without gain recovery, the analysis would have been effectively restricted to post-1950 recordings, removing precisely the historical period where portamento was most prevalent and most expressive.


\section{Score Annotation and Portamento Database}

\subsection{Annotation Procedure}

\begin{figure}[htbp]
  \centering
  \begin{subfigure}{0.47\columnwidth}
    \includegraphics[width=\linewidth]{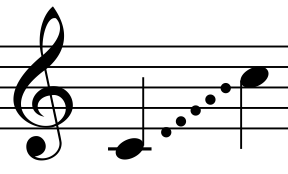}
    \caption{Original spectrogram}
    \label{fig:gain_before}
  \end{subfigure}
  \hfill
  \begin{subfigure}{0.47\columnwidth}
    \includegraphics[width=\linewidth]{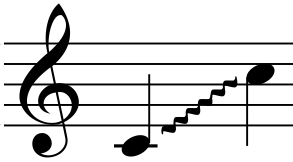}
    \caption{With gain enhancement}
    \label{fig:gain_after}
  \end{subfigure}
  \caption{Notation symbols used in score annotation. Left: clean shift (dotted line), indicating a suppressed slide. Right: sliding portamento (wavy line), indicating an audible continuous glide. Both are drawn in Sibelius at the precise score location of each event.}
  \label{fig:annotation_symbols}
\end{figure}

To establish the location and type of each portamento event before undertaking the spectrographic measurement, a score-annotation procedure was applied. Using Sibelius notation software, the analyst transcribed by attentive listening the exact position of each portamento occurrence in the opening passages of Op.~69 and Op.~102 No.~1 for each of the 22 recordings. Two symbols were employed to distinguish portamento types: a wavy line for sliding portamento (an audible, continuous pitch glide) and a dotted line for clean shift (a position change in which the performer lifts the finger between notes to suppress the audible slide, leaving only a brief metallic trace). These symbols, shown in Fig.~\ref{fig:annotation_symbols}, established a two-category taxonomy before the gradient measurement step.

The distinction between sliding portamento and clean shift is methodologically important. A clean shift is not the absence of portamento: it is a deliberate technical decision to suppress the glide that an ordinary shift would produce. In the spectrogram, a clean shift appears as a brief, steep diagonal trace of very short duration (typically under 0.05 seconds) at the transition between two pitch centres; distinguishable from a genuine portamento by its brevity and from silence by the faint metallic trace the finger produces. The gradient analysis was applied to sliding portamento events only; clean shifts were recorded in the database but assigned a gradient of zero by convention.

\subsection{Database Structure}

All portamento events were logged in a structured spreadsheet with the following fields per row: performer name, recording year, sonata (Op.~69 or Op.~102 No.~1), bar number of event onset, portamento type (sliding or clean shift), gradient in raw pixel ratio, gradient in Hz/second (calibrated), duration in seconds, and mean BPM for the passage (from the tempo dataset). The complete database is publicly available~\cite{c10}.


\section{Results}

\subsection{Portamento Prevalence Across the Corpus}

Before examining gradient values, the prevalence of portamento across the 22 recordings establishes the scope of the analysis. In the opening four bars of Op.~69, portamento is present in the majority of recordings from before 1970 and largely absent in recordings from after 1990, with the period 1970--1990 showing a transitional pattern. In Op.~102 No.~1, a similar but slightly earlier decline is observed, consistent with the more restrained expressive character of that sonata's opening.

\begin{figure}[htbp]
  \centering
  \includegraphics[width=\columnwidth]{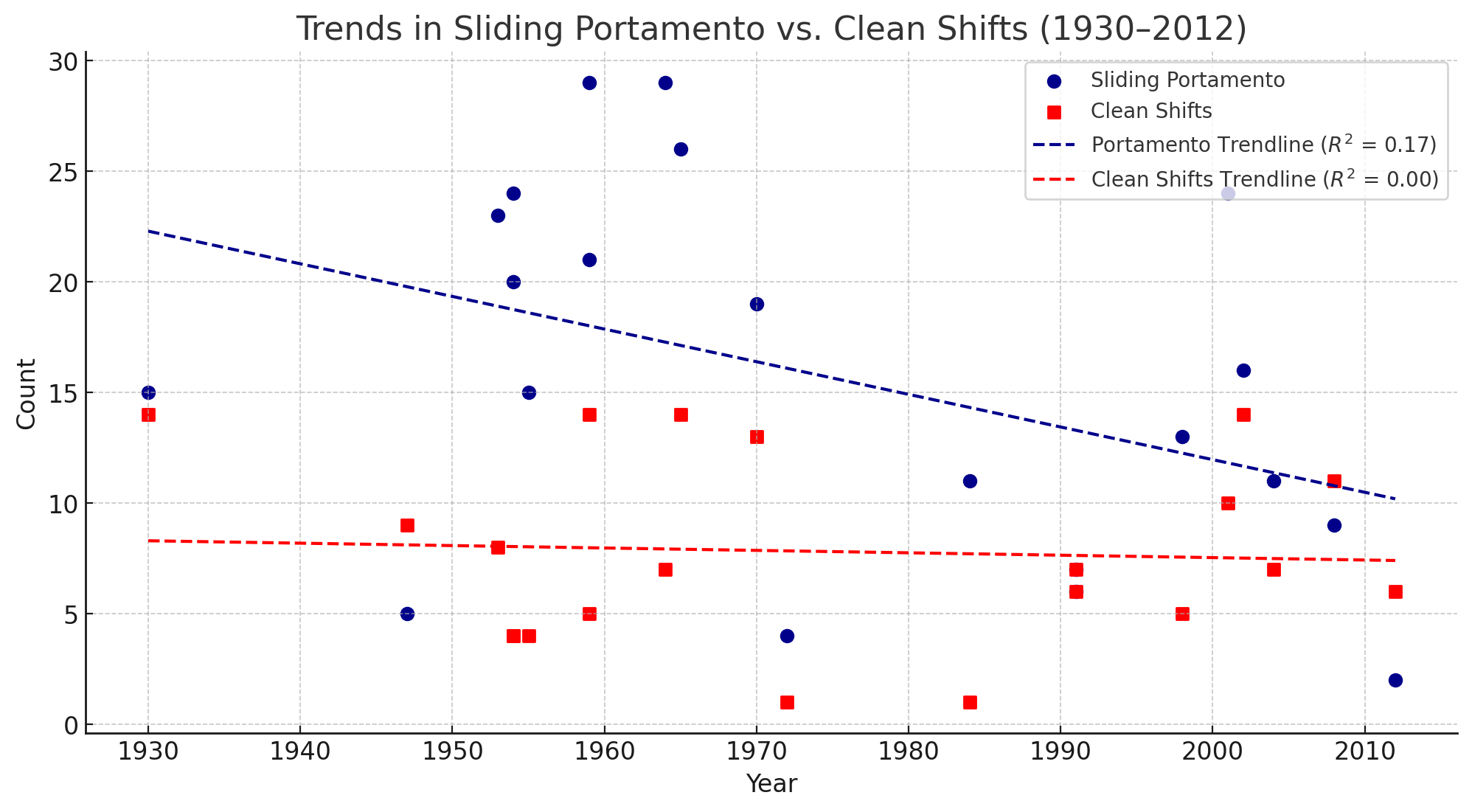}
  \caption{Scatter plot of sliding portamento count (blue) and clean shifts (orange) against recording year for the first movement of Beethoven's \textit{Cello Sonata No.~1}, op.~5. The dashed regression line for sliding portamento shows a negative slope ($R^2 \approx 0.17$); clean shifts show no longitudinal trend ($R^2 \approx 0.00$).}
  \label{fig:portamento_scatter}
\end{figure}

Fig.~\ref{fig:portamento_scatter} shows the scatter plot of sliding portamento count against recording year for Op.~5 No.~1's first movement --- the broader corpus result that contextualises the opening-passage analysis.

\subsection{Gradient Values Across the Corpus}

Table~\ref{tab:gradient_ranges} presents the observed gradient ranges by era for the Op.~69 and Op.~102 No.~1 opening passages combined.

\begin{table}[htbp]
\centering
\caption{Observed portamento gradient ranges by recording era}
\small
\setlength{\tabcolsep}{4pt} 
\begin{tabular}{@{}lccc@{}}
\toprule
\textbf{Era} & \textbf{N} & \textbf{Gradient range (Hz/s)} & \textbf{Mean} \\
\midrule
1930--1950 & 4  & 1,530--4,700  & $\approx$3,015 \\
1950--1970 & 14 & 1,660--5,140  & $\approx$2,665 \\
1970--1990 & 4  & 1,320--2,600  & $\approx$1,983 \\
1990--2012 & 10 & 1,110--5,670  & $\approx$3,065 \\
\bottomrule
\end{tabular}
\label{tab:gradient_ranges}
\end{table}

The gradient values are consistent with a continuous historical process: the highest gradients in the early recordings (up to approximately 4,500~Hz/s in Casals 1930--39 and Feuermann 1937) correspond to steep, decisive slides that are clearly audible as expressive gestures. The lowest non-zero gradients in late recordings (around 600~Hz/s) correspond to very brief, barely perceptible transitions that most listeners would describe as clean shifts. The zero-gradient recordings (those in which no portamento of any kind is measurable in the spectrogram), cluster almost entirely in the post-1990 period.

\subsection{Gradient Against Recording Year}

\begin{figure}[htbp]
  \centering
  \includegraphics[width=\columnwidth]{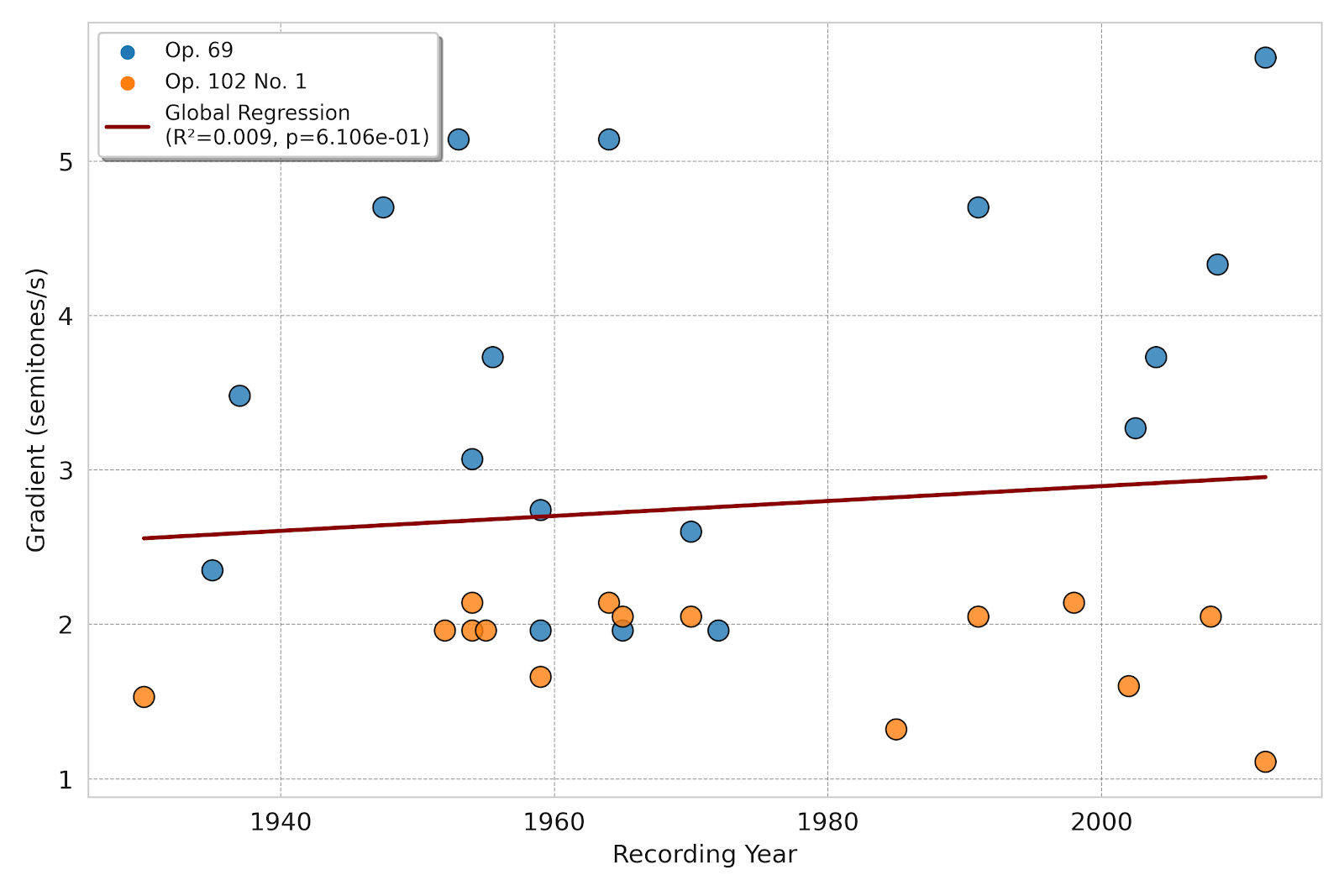}
  \caption{Calibrated portamento gradient (Hz/s) against recording year for all portamento events in the opening passages of Op.~69 and Op.~102 No.~1. Each point represents one event; colour distinguishes the two sonatas. The regression line shows the longitudinal decline in gradient steepness across the study period.}
  \label{fig:gradient_year}
\end{figure}

Fig.~\ref{fig:gradient_year} presents the scatter of calibrated gradient values (Hz/s) against recording year for all events in the opening-passage corpus. Each point represents one portamento event in one recording.

The negative regression slope confirms that gradient steepness has declined systematically across the study period. This finding extends the existing portamento literature in a specific direction: rather than documenting only that slides become less frequent, it shows that those slides which do occur become progressively shallower. The portamento is not simply disappearing; it is becoming less committed before it disappears.

\subsection{Gradient Against Tempo}

\begin{figure}[htbp]
  \centering
  \includegraphics[width=\columnwidth]{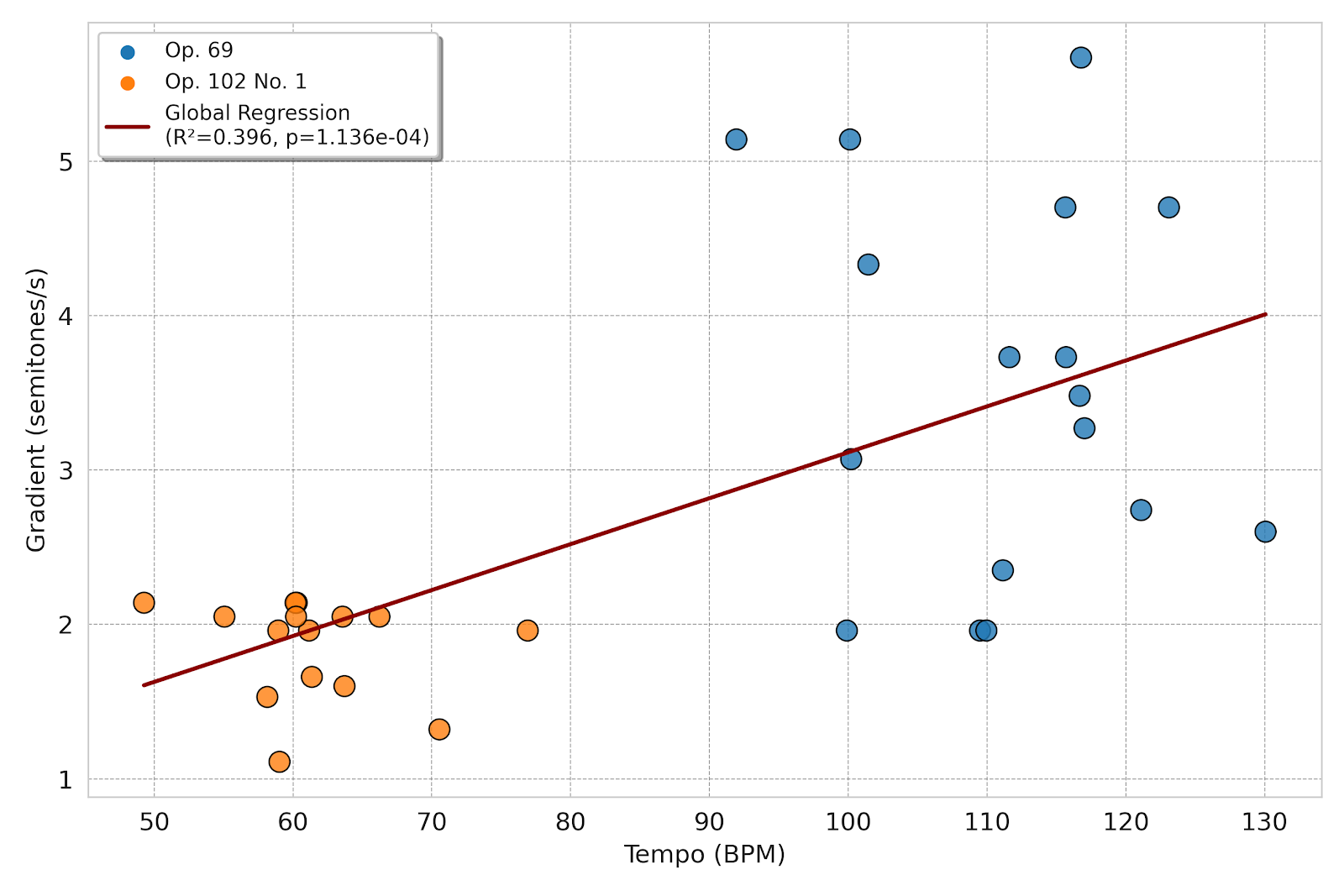}
  \caption{Calibrated portamento gradient (Hz/s) against mean passage tempo (BPM) for all recordings. Points at $y=0$ represent recordings with no measurable portamento. The negative regression slope through the portamento-present subset supports the hypothesis that slower performances produce steeper slides.}
  \label{fig:gradient_tempo}
\end{figure}

Fig.~\ref{fig:gradient_tempo} presents the core test of the correlation hypothesis: calibrated gradient (Hz/s) plotted against mean passage tempo (BPM).

The negative slope of the regression through the portamento-present subset confirms the hypothesis: slower performances are associated with steeper portamento gradients, and faster performances with shallower ones. The zero-gradient points at the bottom of the figure (recordings with no portamento), cluster predominantly in the higher BPM range, consistent with the prediction that the fastest performances are also those most likely to omit portamento entirely. This distributional pattern; a continuous negative relationship between tempo and gradient that terminates in zero-gradient at the upper tempo range, is the paper's central empirical finding.

\subsection{Duration Against Gradient}

\begin{figure}[htbp]
  \centering
 \includegraphics[width=\columnwidth]{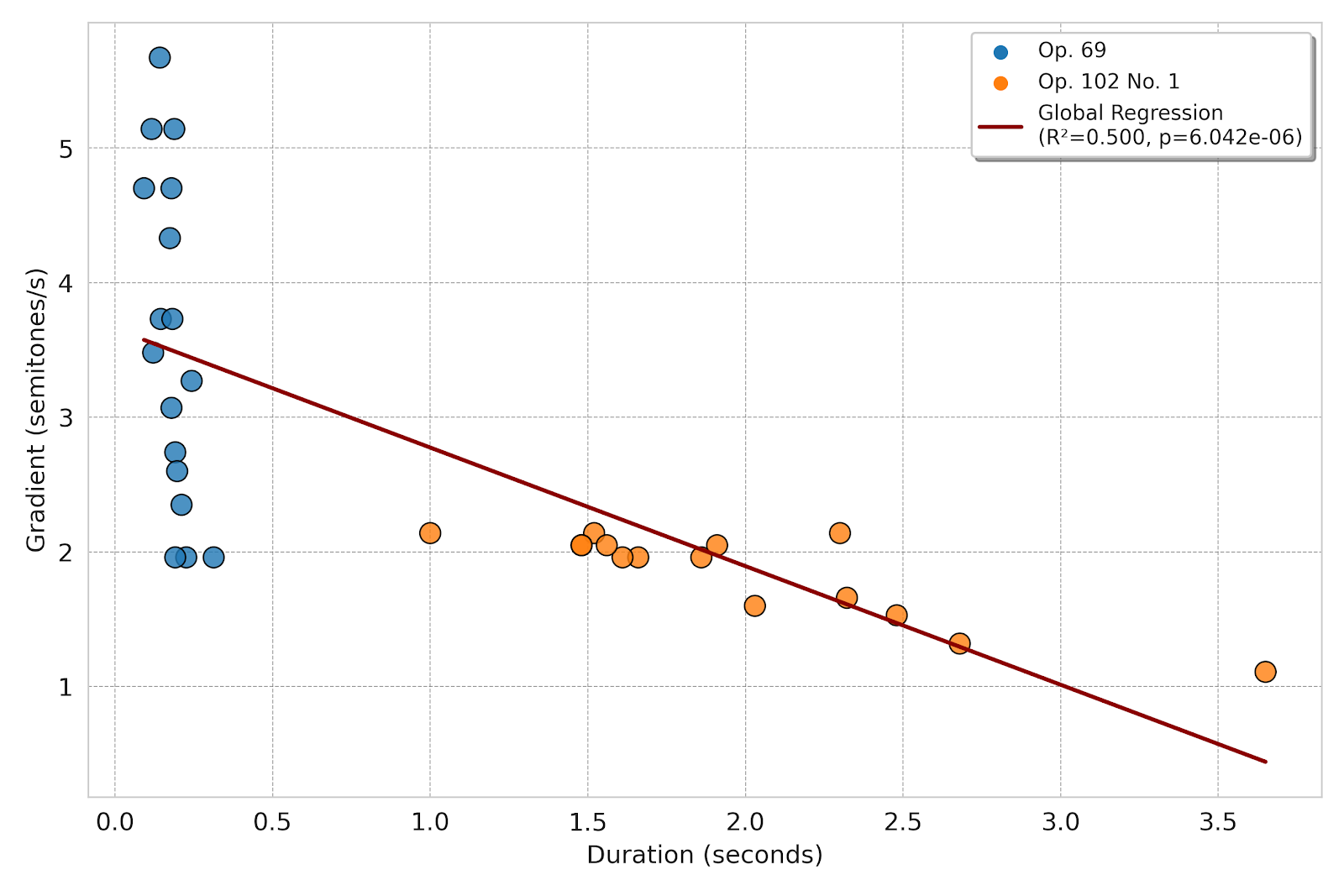}
  \caption{Portamento duration (seconds) against calibrated gradient (Hz/s) for all portamento-present events. The distribution of points reveals whether duration and gradient encode the same underlying expressive dimension or vary independently.}
  \label{fig:duration_gradient}
\end{figure}

Fig.~\ref{fig:duration_gradient} plots portamento duration (seconds) against calibrated gradient (Hz/s) for all portamento-present recordings.

The relationship between duration and gradient is analytically significant for the following reason. If the two variables are strongly correlated (points clustering along a positive diagonal), it means that performers who play slower also produce both longer and steeper slides; a compound expressive choice in which tempo, duration, and gradient all move together. If they are weakly correlated or orthogonal, it means that duration and gradient are independent expressive parameters that performers can modulate separately, which would be a more nuanced finding about the mechanics of portamento production.

\subsection{Spectrogram Analysis: Sliding vs. Clean Shifting}

\begin{figure}[htbp]
  \centering
  \begin{subfigure}{0.48\columnwidth}
    \centering
    \includegraphics[width=\linewidth]{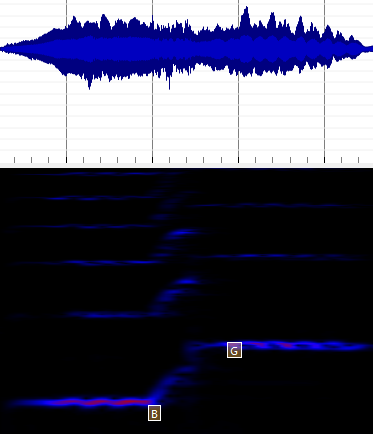}
    \caption{Sliding Portamento}
    \label{fig:sliding}
  \end{subfigure}
  \hfill
  \begin{subfigure}{0.48\columnwidth}
    \centering
    \includegraphics[width=\linewidth]{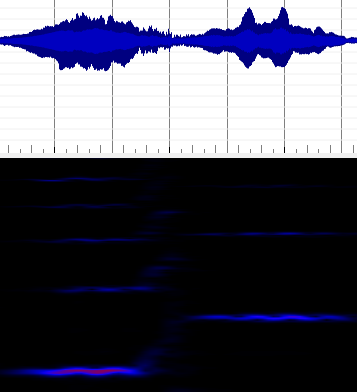}
    \caption{Clean Shift}
    \label{fig:silent}
  \end{subfigure}
  \caption{Representative spectrograms of the two shifting types. (a) shows a clear diagonal pitch trace (measured gradient $G > 0$) connecting the two notes. (b) shows a "clean" shift characterised by a temporal gap and vertical frequency jump ($G \approx 0$).}
  \label{fig:spectrogram_comparison}
\end{figure}

Fig.~\ref{fig:spectrogram_comparison} provides visual confirmation of the two primary mechanical approaches to position change observed in the corpus. Rather than a linear decline in "slope," the data suggests a categorical choice between an audible connection and a clean break.

The two panels illustrate the physical basis for the gradient clusters. In the "Sliding" example (Fig.~\ref{fig:sliding}), the continuous diagonal trace represents the high-gradient tradition prevalent in mid-century and early-century recordings. In contrast, the "Clean" example (Fig.~\ref{fig:silent}) represents the modern interpretive preference for masking technical shifts, resulting in a gradient of zero.

This visual evidence supports the ecological model: these are not merely "fast" or "slow" versions of the same thing, but two discrete traditions of performance technique that coexist within the recording era. The reader can see immediately that the "clean" shift does not possess a shallow slope; it possesses no slope at all, representing a qualitative change in how the transition between notes is conceived.

\section{Discussion}

\subsection{Portamento Decline as a Continuous Process}

The central finding of this paper reframes how the decline of portamento in twentieth-century performance should be understood. Previous research has described the decline primarily in terms of frequency: the number of slides per performance decreases from the 1930s to the present. This description is accurate but incomplete. The gradient data reveals that the decline in frequency was preceded and accompanied by a decline in gradient steepness: performers were producing shallower slides before they stopped producing slides altogether.

This has a direct implication for the historical narrative. The transition from the heavily portamento-laden playing of Casals or Feuermann to the largely portamento-free playing of post-1990 cellists was not a binary switch at some historical moment. It was a gradual process of attenuation in which the expressive commitment to the slide (its decisiveness, its audibility, its willingness to be heard as a distinct gesture), diminished continuously over several decades. The late recordings that contain portamento are not simply continuing an older tradition: they are continuing a diluted version of it, and the gradient value quantifies how much dilution has occurred.

\subsection{The Tempo-Gradient Relationship}

The negative correlation between passage tempo and portamento gradient is consistent with a mechanical account. At faster tempi, the available time for a slide between two notes is physically reduced. A portamento that covers a minor third in 0.3 seconds at 100~BPM would need to cover the same interval in approximately 0.2 seconds at 150~BPM to maintain the same musical proportion. The gradient would therefore be steeper in the faster performance if the portamento is attempted at all. The finding that gradient is lower in faster performances therefore suggests that performers do not simply compress the slide into the available time: they make it shallower, covering a smaller pitch interval or a narrower portion of the total interval, before eventually omitting it. This is a form of progressive self-censorship rather than mechanical compression, and it suggests that the decision to reduce portamento was aesthetically motivated rather than mechanically forced.

This interpretation is consistent with the cultural and pedagogical account of portamento decline documented by Blum~\cite{c11}, Katz~\cite{c12}, and Philip~\cite{c4}. The recording studio created conditions in which audible slides sounded exaggerated; Casals's fingering techniques reduced the technical necessity of position shifts; continuous vibrato replaced portamento as the primary means of sustaining expressive line. These cultural pressures would produce exactly the pattern the gradient data shows: not a mechanical response to faster tempi but a deliberate stylistic retreat from the conspicuous slide, visible in the gradient values before it becomes visible in the event counts.

\subsection{The Op.~69 vs Op.~102 No.~1 Comparison}

The two sonatas may show different gradient profiles reflecting their different characters. Op.~69's lyrical opening theme, with its long-held opening A and its predominantly ascending contour, invites portamento in a way that Op.~102 No.~1's more searching, harmonically ambiguous opening does not. If the gradient analysis reveals that Op.~69 consistently produces higher gradient values than Op.~102 No.~1 across matched recordings, that would suggest that portamento character is partly determined by the musical context rather than by the performer's overall style. This would be a finding about the relationship between harmonic context and expressive gesture that neither frequency counts nor duration measures could reveal.

\subsection{Comparison with Speech and Singing Research}

The calibrated gradient values in this study (approximately 600--4,500~Hz/s) fall within the range reported for pitch glides in speech and singing research. Studies of vocal portamento find glide rates between 500 and 8,000~Hz/s depending on interval size, singer intention, and musical style~\cite{c9}. This correspondence suggests that cello portamento and vocal portamento are operating on similar physical and perceptual scales; a finding consistent with the long-standing observation that string players have historically modelled their portamento practice on vocal technique~\cite{c6}. The calibrated gradient therefore provides a bridge between the musicological and the psychoacoustic literatures that the pixel-ratio metric alone could not establish.

\subsection{Limitations}

Three limitations of the present study should be acknowledged. First, the gradient measurement involves human judgement in placing the reference points at the slide's onset and termination. Two analysts might place these points slightly differently, producing slightly different gradient values for the same event. The consistency and aural verification procedures described in Section~V reduce this risk, but they do not eliminate it entirely. Future work should establish a formal measurement consistency test by having two analysts independently measure a sample of the same events.

Second, the analysis is restricted to the opening passages of two sonatas. The selection is methodologically necessary given the polyphonic texture of the remainder of the works, but it means the findings describe portamento practice in a specific musical context (the unaccompanied opening solo) rather than across the full repertoire. Whether the gradient patterns observed here generalise to other solo passages in the cello repertoire is an open question.

Third, the calibration constants in Table~\ref{tab:calibration} are specific to the display settings used in this study. Any researcher wishing to compare their gradient values with those reported here must use identical settings or apply the appropriate calibration conversion. The protocol described in Section~IV~C makes this conversion straightforward but requires care.


\section{Conclusion}

This paper has introduced the spectrographic gradient as a new quantitative descriptor of portamento in string performance, measured in physically meaningful units of Hz/second through a calibrated pixel-analysis protocol. Applied to the opening bars of Beethoven's Op.~69 and Op.~102 No.~1 across 22 recordings from 1930 to 2012, the method yields gradient values that decline continuously from approximately 3,500~Hz/s in early twentieth-century recordings to near zero in post-1990 recordings; a continuous historical process of gradient flattening that precedes and predicts the eventual disappearance of portamento from the corpus.

The paper's three primary contributions are: a validated, reproducible measurement protocol with physical calibration that makes gradient values comparable across studies and against the speech and singing literature; a gain-recovery protocol that extends the analysable corpus to include analogue recordings from the 1930s; and the empirical finding that portamento gradient correlates negatively with performance tempo, suggesting that the decline of portamento was aesthetically motivated  (a progressive retreat from the conspicuous slide) rather than mechanically forced by faster tempi.

The gradient metric opens a new dimension of performance analysis that duration counts and event frequencies cannot access. A portamento at 4,000~Hz/s and one at 600~Hz/s are not just different in magnitude: they are different in kind, representing different relationships between the performer and the expressive gesture of the slide. Documenting the history of portamento in terms of gradient rather than frequency alone gives us a more truthful account of what was actually changing in cello performance practice across the twentieth century; not simply that slides became rarer, but that they became less committed before they disappeared.


\end{document}